\documentclass{article}
\usepackage[a4paper, total={6in, 8in}, top = 1in, bottom = 1in]{geometry}
\usepackage{graphicx}
\usepackage{graphics}
\usepackage{hyperref}
\usepackage{lineno}
\usepackage{xspace}
\usepackage{amssymb}
\usepackage{amsmath}
\usepackage{mathrsfs}
\usepackage{epstopdf}
\newcommand{\PbPb}         {\mbox{Pb--Pb}\xspace}

\newcommand{\pPb}          {\mbox{p--Pb}\xspace}

\newcommand{\pt}           {\ensuremath{p_{\rm T}}\xspace}

\newcommand{\nineH}        {$\sqrt{s}~=~0.9$~Te\kern-.1emV\xspace}
\newcommand{\seven}        {$\sqrt{s}~=~7$~Te\kern-.1emV\xspace}
\newcommand{\twoH}         {$\sqrt{s}~=~0.2$~Te\kern-.1emV\xspace}
\newcommand{\twosevensix}  {$\sqrt{s}~=~2.76$~Te\kern-.1emV\xspace}
\newcommand{\five}         {$\sqrt{s}~=~5.02$~Te\kern-.1emV\xspace}
\newcommand{\twosevensixnn}{$\sqrt{s_{\mathrm{NN}}}~=~2.76$~Te\kern-.1emV\xspace}
\newcommand{\fivenn}       {$\sqrt{s_{\mathrm{NN}}}~=~5.02$~Te\kern-.1emV\xspace}

\newcommand{\MeVc}         {Me\kern-.1emV/$c$\xspace}

\newcommand{\GeV}          {Ge\kern-.1emV\xspace}
\newcommand{\MeV}          {Me\kern-.1emV\xspace}
\newcommand{\GeVmass}      {Ge\kern-.2emV/$c^2$\xspace}
\newcommand{\MeVmass}      {Me\kern-.2emV/$c^2$\xspace}

\newcommand{\ee}           {\ensuremath{\mathrm{e^{+}e^{-}}}\xspace}

\newcommand {\meanpT}    {\ensuremath{\langle p_{\mathrm{T}} \kern-0.1em\rangle}\xspace}
\newcommand {\mean}[1]   {\ensuremath{\langle #1 \kern-0.1em\rangle}\xspace}

\newcommand {\ep}        {\mbox{$\mathrm {e^-p}$}\xspace}

\newcommand {\MeanNpart} {\mbox{\ensuremath{< \kern-0.15em N_{part} \kern-0.15em >}}}

\newcommand {\mass}     {\mbox{\rm MeV$\kern-0.15em /\kern-0.12em c^2$}}
\newcommand {\tev}      {\textrm{TeV}\xspace}
\newcommand{\TeV} {\tev}

\newcommand {\mmom}     {\mbox{\rm MeV$\kern-0.15em /\kern-0.12em c$}}
\newcommand {\gmom}     {\mbox{\rm GeV$\kern-0.15em /\kern-0.12em c$}}
\newcommand {\mmass}    {\mbox{\rm MeV$\kern-0.15em /\kern-0.12em c^2$}}
\newcommand {\gmass}    {\mbox{\rm GeV$\kern-0.15em /\kern-0.12em c^2$}}

\newcommand {\dg}       {\mbox{$\kern+0.1em ^\circ$}}

\newcommand{\gevc}{\ensuremath{\mathrm{GeV}/c}\xspace}
\newcommand{\GeVc}{\gevc}

\newcommand{\lambdac}     {\ensuremath{\mathrm {\Lambda_{c}^{+}}}\xspace}
\newcommand{\xicz}        {\ensuremath{\mathrm {\Xi_{c}^{0}}}\xspace}
\newcommand{\xiczp}        {\ensuremath{\mathrm {\Xi_{c}^{0,+}}}\xspace}
\newcommand{\XicD} {\ensuremath{\xiczp/\Dz}\xspace}
\newcommand{\xicp}        {\ensuremath{\mathrm {\Xi_{c}^{+}}}\xspace}

\newcommand{\rmLambdas}         {\ensuremath{\mathrm {\Lambda \kern-0.2em + \kern-0.2em \overline{\Lambda}}}\xspace}

\newcommand{\Kzs}               {\ensuremath{\mathrm {K^0_S}}\xspace}

\newcommand{\Dzero}{\ensuremath{\mathrm {D^0}}\xspace}
\newcommand{\Dz}{\Dzero}

\newcommand{\Dplus}{\ensuremath{\rm D^+}\xspace}

\newcommand{\Dsubs}{\ensuremath{\rm D_{s}^+}\xspace}
\newcommand{\Ds}{\Dsubs}

\newcommand{\Lcplus}{\lambdac}
\newcommand{\Lc}         {\Lcplus}
\newcommand{\LcD} {\ensuremath{\lambdac/\Dzero}\xspace}

\newcommand{\sqrtsfive}{\ensuremath{\sqrt{s} = 5.02~\TeV}\xspace}
\newcommand{\sqrtsthirt}{\ensuremath{\sqrt{s} = 13~\TeV}\xspace}

\newcommand{\sqrtsNNfive}{\ensuremath{\sqrt{s_\mathrm{NN}} = 5.02~\TeV}\xspace}

\newcommand{\figref}[1]{Fig.~\ref{#1}}

\newcommand{\Omegac}{\ensuremath{\Omega_{\rm c}^{0}}\xspace}
\newcommand{\Sigmac}{\ensuremath{\Sigma_{\rm c}^{0,++}}\xspace}

\newcommand{\lowptbin}{\ensuremath{0<\pt<1~\GeVc}\xspace}
\usepackage{subcaption}

\title{\textbf{Charm production and hadronization in pp and p--Pb collisions at the LHC with ALICE}}

\author{Annalena Kalteyer\\
on behalf of the ALICE Collaboration\\ 
GSI Helmholtzzentrum für Schwerionenforschung GmbH\\
Planckstraße 1, 64291 Darmstadt\\
Heidelberg University}

\date{}
\usepackage{color}
\usepackage{lineno}

\begin{document}
% \begin{titlepage}

\maketitle
\thispagestyle{empty}

\vspace{0.5cm}
Presented at DIS2022: XXIX International Workshop on Deep-Inelastic Scattering and Related Subjects, Santiago de Compostela, Spain, May 2-6 2022.
\vspace{0.5cm}

\begin{abstract}
%\linenumbers
Studies of open-charm hadron production in pp and \pPb collisions are performed by the ALICE Collaboration at the LHC to investigate charm-quark hadronization mechanisms. Recent measurements of charm meson (${\rm D^0}$, ${\rm D^+}$, ${\rm D^+_{\rm s}}$, and ${\rm D^{*+}}$) and baryon ($\Lambda^+_{\rm c}$, $\Xi^{0,+}_{\rm c}$, $\Sigma^{0,++}_{\rm c}$, and $\Omega^0_{\rm c}$) production in pp collisions at $\sqrt{s}$ = 5.02 TeV and $\sqrt{s}$ = 13 TeV allow the determination of the charm cross section and the  fragmentation fractions at $\sqrt{s}$ = 5.02 TeV with unprecedented precision. The measurements show that the fragmentation fractions significantly differ from the ones measured in ${\rm e^+e^-}$ and \ep collisions. This highlights possible differences in the hadronization process between electron induced collisions and pp collisions at the LHC. \\
Furthermore, the first measurement of the baryon-to-meson yield ratio $\Lambda^+_{\rm c}/{\rm D^0}$, down to $p_{\rm T}=0$ in pp and \pPb collisions will be discussed. In p--Pb collisions a modification of the hadronization mechanisms could be present due to cold nuclear matter effects and collective phenomena. A systematic comparison between data and model calculations will help to understand charm quark hadronization in pp and p--Pb collisions.
\end{abstract}

\vspace{0.5cm}

% \end{titlepage}
% \setcounter{page}{2} %please do not remove this line

% \mbox{}
% \vfill

\section{Introduction} 
\begin{figure}[h!b]
    \centering
    \includegraphics[width=0.45\textwidth]{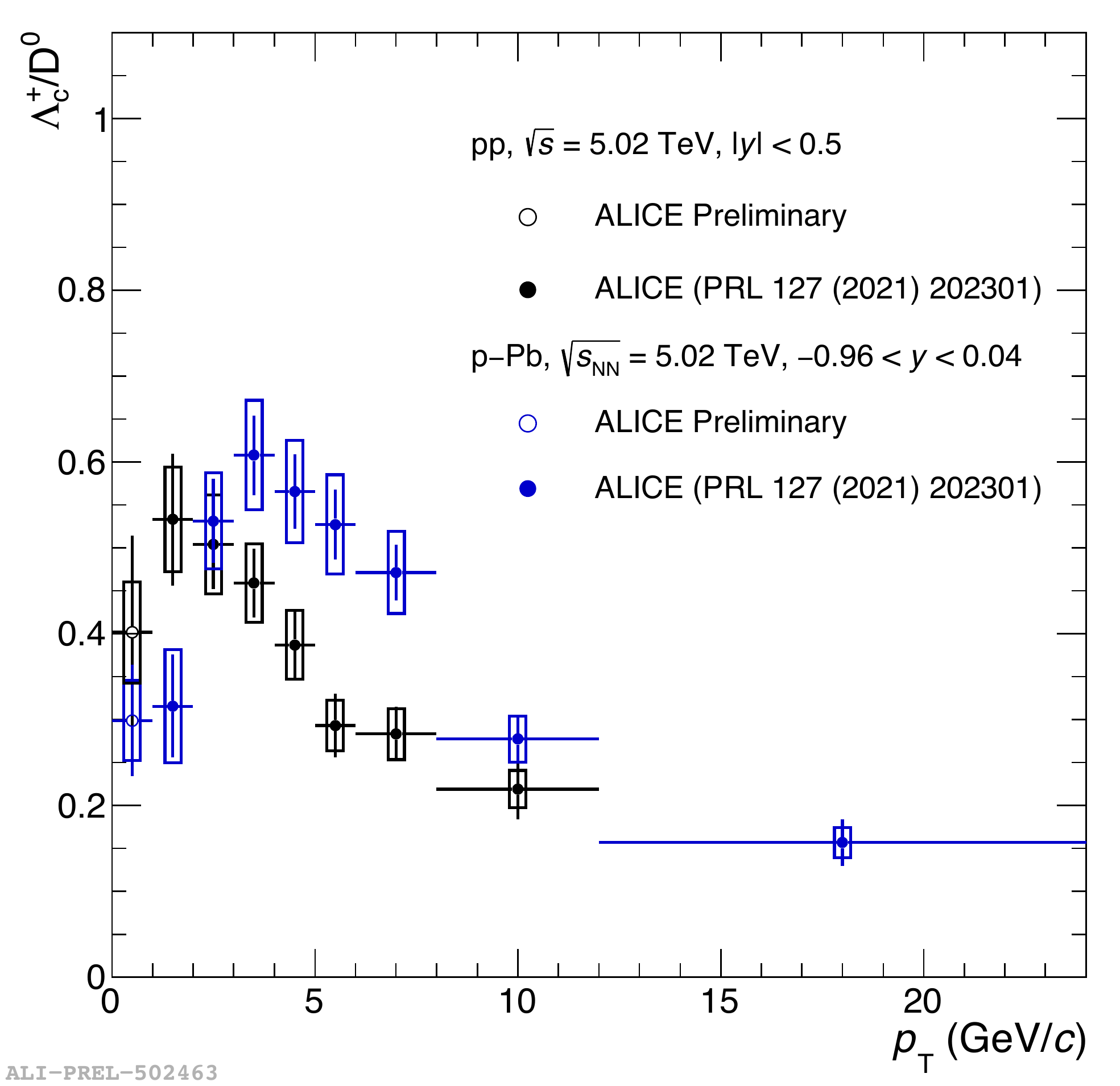}
    \includegraphics[width=0.45\textwidth]{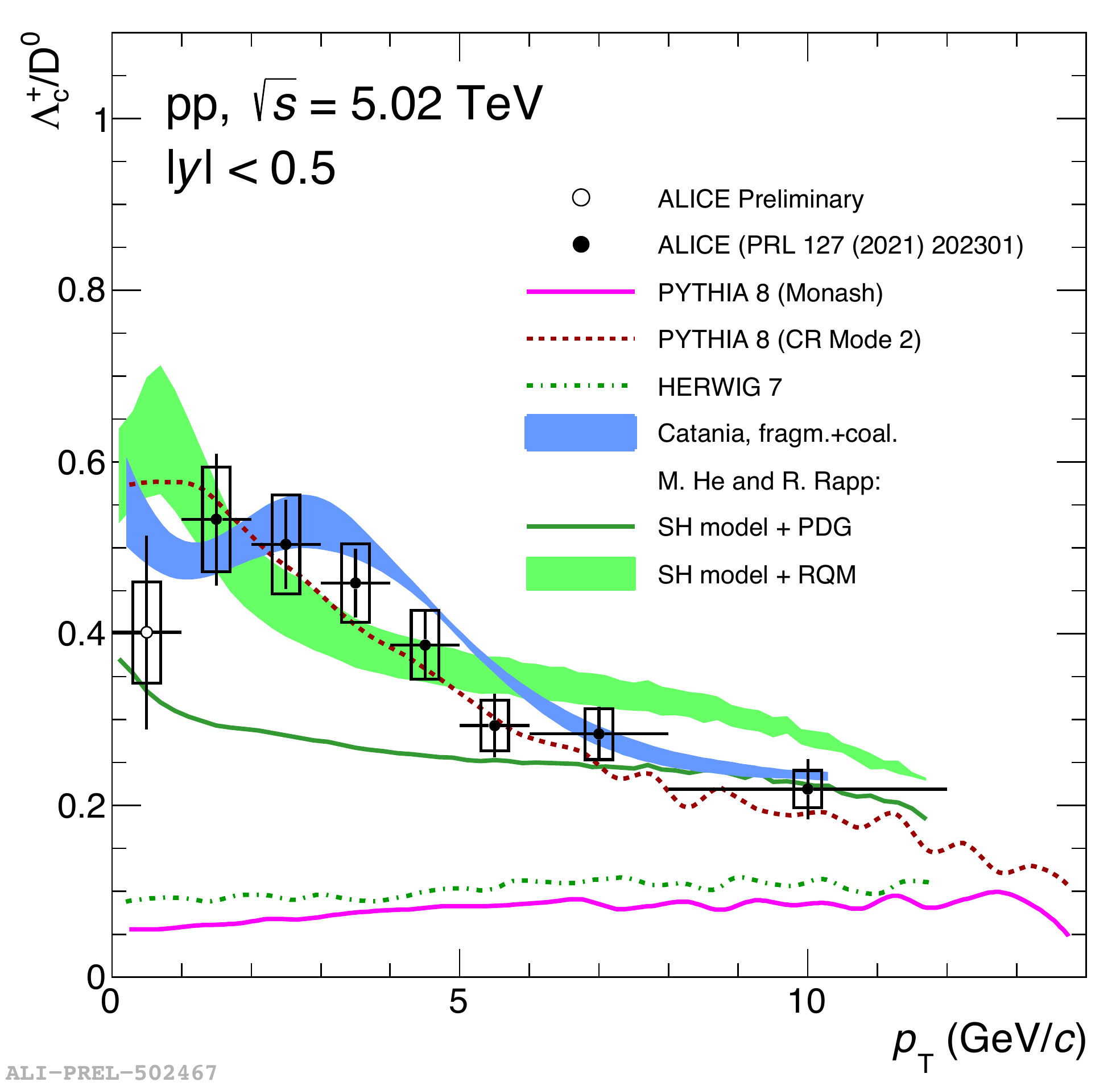}
    \includegraphics[width=0.45\textwidth]{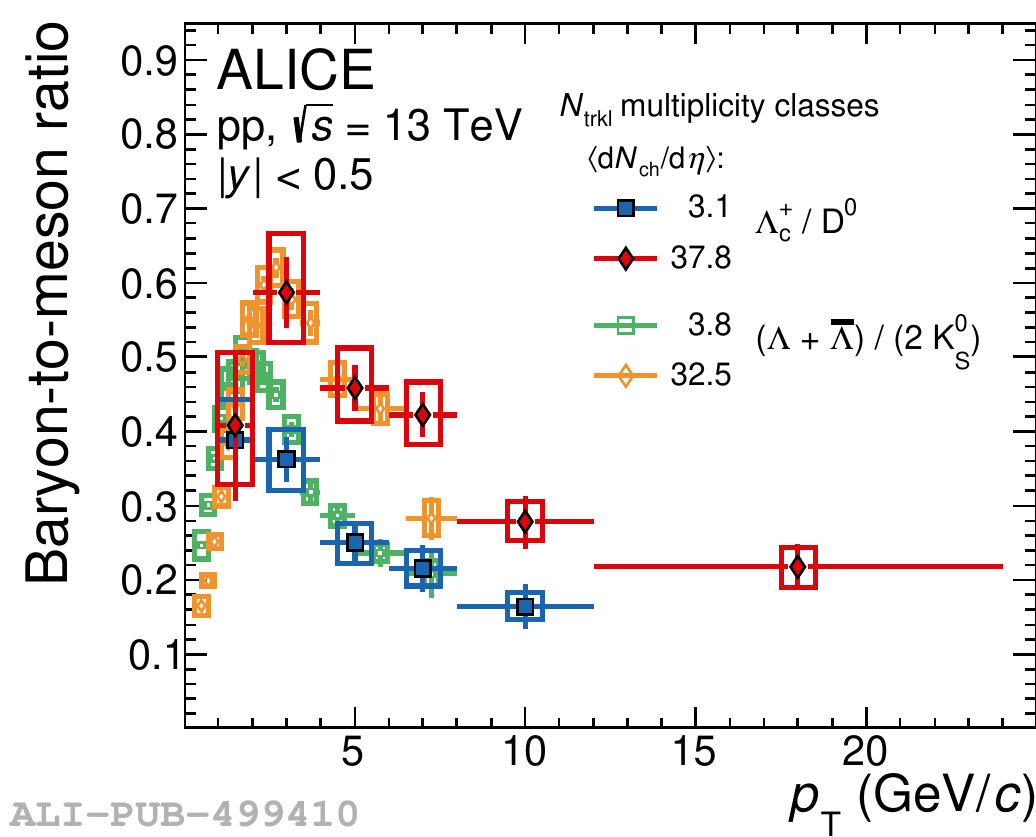} 
    \includegraphics[width=0.45\textwidth]{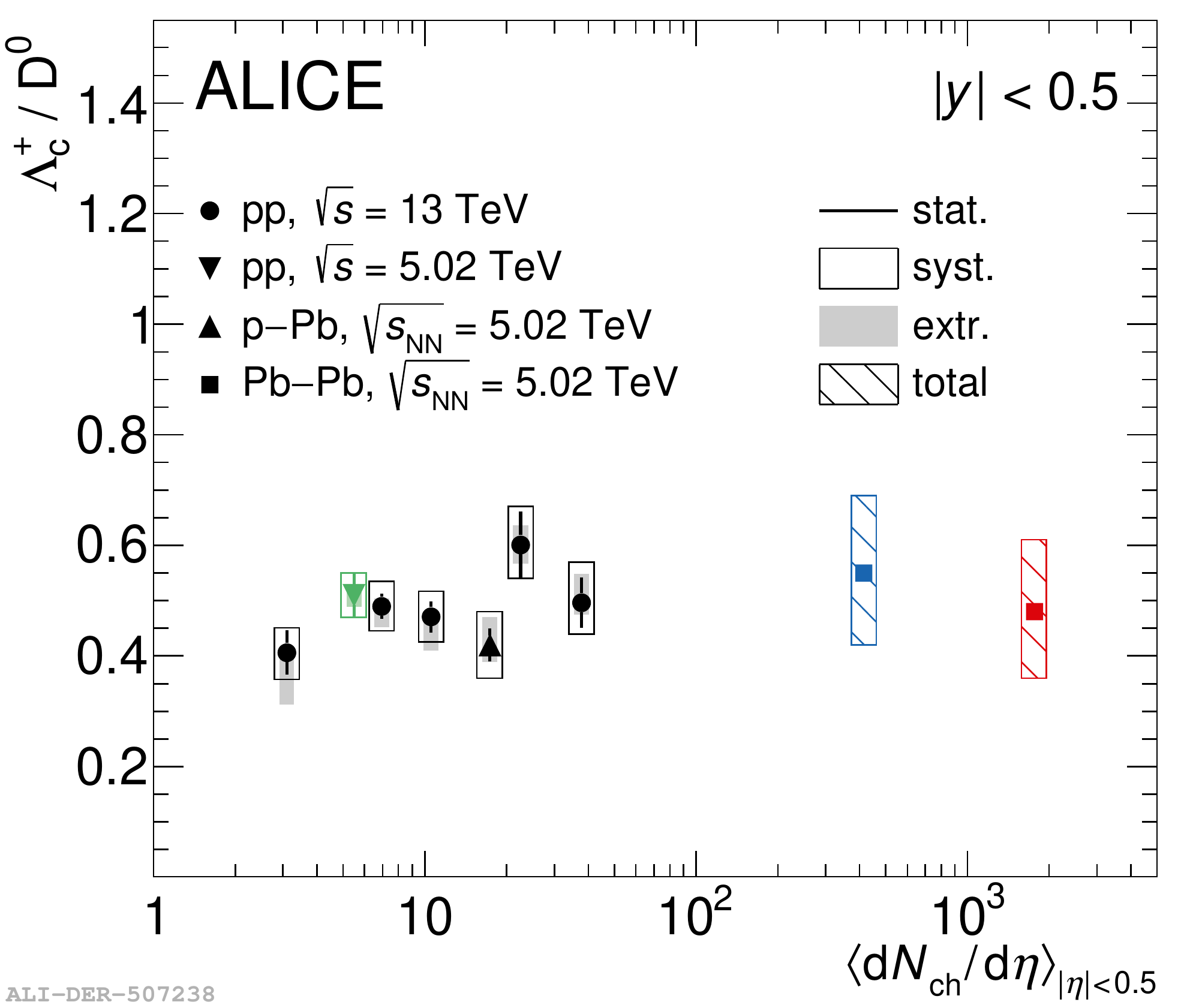} 
    \caption{Upper left: \LcD ratio in pp and \pPb collisions at \sqrtsNNfive as a function of \pt~\cite{ALICE:2020wla}. Upper right: \LcD ratio as a function of \pt in pp collisions at \sqrtsfive, including comparisons to models~\cite{He:2019tik,Christiansen:2015yqa,Skands:2014pea,Minissale:2020bif,Li:2021nhq}. Lower left: \LcD and $\Lambda/\mathrm{K_s}^0$ ratio in pp collisions at \sqrtsthirt in different multiplicity classes~\cite{ALICE:BaryonMesonMult},~\cite{ALICE:2013wgn}.  Lower right: \pt-integrated \LcD yield ratio as a function of charged-particle multiplicity in pp, \pPb and \PbPb collisions in ALICE, measured at various collision energies~\cite{acharya2022observation}.} 
    \label{fig:LcDWithModels}  \label{fig:LcD}
\end{figure}

 Charm and beauty quarks have masses much larger than the $\Lambda_\mathrm{QCD}$ energy scale, which means that their production can be calculated perturbatively. Thus, measurements of heavy-flavor hadron production provide crucial tests for calculations of quantum chromodynamics (QCD). Typically, heavy-flavor hadron production cross sections are calculated in a factorization approach multiplying three separate components: the parton distributions functions (PDFs), which describe the momentum distributions of quarks and gluons within the incoming hadrons; the hard-scattering cross section for the partons to produce a charm or beauty quark; and the fragmentation functions, which characterize the hadronization of a quark to its respective hadron species. As the fragmentation functions are non-perturbative, they are determined from measurements in \ee and \ep collisions, and the relevant hadronization processes were so far assumed to apply universally, independent of the energy and collision system. In particular, ratios of heavy-flavour hadrons are sensitive to the charm fragmentation functions, since the terms related to the PDFs and partonic cross sections are common to all hadrons and cancel out in the yield ratios. Recent measurements performed by the ALICE Collaboration of the charm meson-to-meson yield ratios $\Dplus/\Dz$ and $\Ds/\Dz$~\cite{ALICE:2021mgk} have shown that there is no momentum dependent modification of the charm meson hadronization, since the measurements are flat as a function of \pt. The results are also in agreement with measurements performed at \ee and \ep colliders, and with pQCD calculations based on the factorization approach that include fragmentation functions parameterized from \ee and \ep collisions. On the contrary, the charm baryon-to-meson yield ratio \LcD, measured in pp collisions at \sqrtsfive, is strongly \pt dependent~\cite{ALICE:2020wla}, as shown in the upper right panel of \figref{fig:LcD}. The \LcD yield ratio is not described by MC generators, such as PYTHIA with Monash tune~\cite{Skands:2014pea} and HERWIG 7~\cite{bellm2016herwig}, which use fragmentation functions parameterized from \ee collisions. However, various models do describe the measured baryon-to-meson ratio. For instance PYTHIA 8 including color reconnection (CR) beyond leading color~\cite{Christiansen:2015yqa} enhances the baryon production in multi-parton interactions by introducing color junctions. The Catania model~\cite{Minissale:2020bif} considers a thermalized systems of light quarks, anti-quarks, and gluons and treats hadronization as a combination of fragmentation and coalescence. The statistical hadronization model, which is based on fragmentation functions from \ee collisions, includes unobserved states in addition to those listed by the PDG~\cite{particle2020review} by taking guidance from the relativistic quark model (RQM)~\cite{He:2019tik}. \\
 The enhancement with respect to electron collider measurements, and the strong \pt dependence for the \LcD yield ratio is also observed for \XicD, $\Omegac/\Dz$, and $\Sigmac/\Dz$ yield ratios in the measured momentum range. In addition, more differential measurements were performed, like measuring the \LcD yield ratio as a function of the event multiplicity. The enhancement of the \LcD yield ratio even persists in low multiplicity pp collisions as it will be discussed in the next section.
 
 \section{Results}
 \subsection{\boldmath{\LcD} yield ratio}

 In the upper left panel of \figref{fig:LcD} the baryon-to-meson yield ratio \LcD is presented down to $\pt=0$ in pp and \pPb collisions at \sqrtsNNfive. A significant fraction of the total hadron cross section is at low momenta. Thus, an extension of the measurement range from the published results at \pt $>1$ GeV/c~\cite{ALICE:2020wfu}, shown with the full markers, to the preliminary results in the interval \lowptbin, indicated by the open circles, is crucial to complete the picture of hadronization. The low momentum range is dominated by combinatorial background, but a new candidate reconstruction method using the KFParticle package~\cite{kfparticle} and applying machine-learning selections with the XGBoost gradient boosting algorithm~\cite{chen2016xgboost} to separate signal and background, allows the signal extraction down to $\pt=0$. The \pPb result at \pt~$>4$ GeV/c hints at a larger \LcD yield ratio than the one measured in pp collisions, while the ratio is lower at \pt~$<4$ GeV/c. This suggests a possible \pt redistribution of baryons in \pPb collisions with respect to pp collisions. This effect could be attributed to a contribution of collective effects, i.e. radial flow, in \pPb collisions, which is consistent with previous observations in the light-flavor $\Lambda/\Kzs$ baryon-to-meson yield ratio~\cite{ALICE:2013wgn}. \\
 A multiplicity differential measurement in pp collisions at \sqrtsthirt of the \LcD yield ratio as a function of the transverse momentum~\cite{ALICE:BaryonMesonMult} is shown in the lower left panel of \figref{fig:LcD}. The figure shows that there is an increase in the yield ratio from low to high multiplicity for $1< \pt < 12$~GeV/c, which could also point to a different momentum redistribution among baryons and mesons. A similar effect was also seen in the strange sector when looking at the $\Lambda/\Kzs$ ratio, also shown in the lower left panel of \figref{fig:LcD}. This points to a similar hadronization mechanism in the light and heavy flavor sectors. However, when studying the \pt integrated \LcD yield ratio as a function of multiplicity~\cite{acharya2022observation} as shown in the lower right panel of \figref{fig:LcD} there seems to be no significant variation as a function of the multiplicity, collision system or energy. This suggests that the modification observed in the \pt differential studies could be due to a momentum redistribution without a modification of the overall yield. 
 
 \subsection{\boldmath{\Sigmac} production in pp collisions at \boldmath{\sqrtsthirt}}
\begin{figure}[h!tb]
    \centering
    \includegraphics[width=\textwidth]{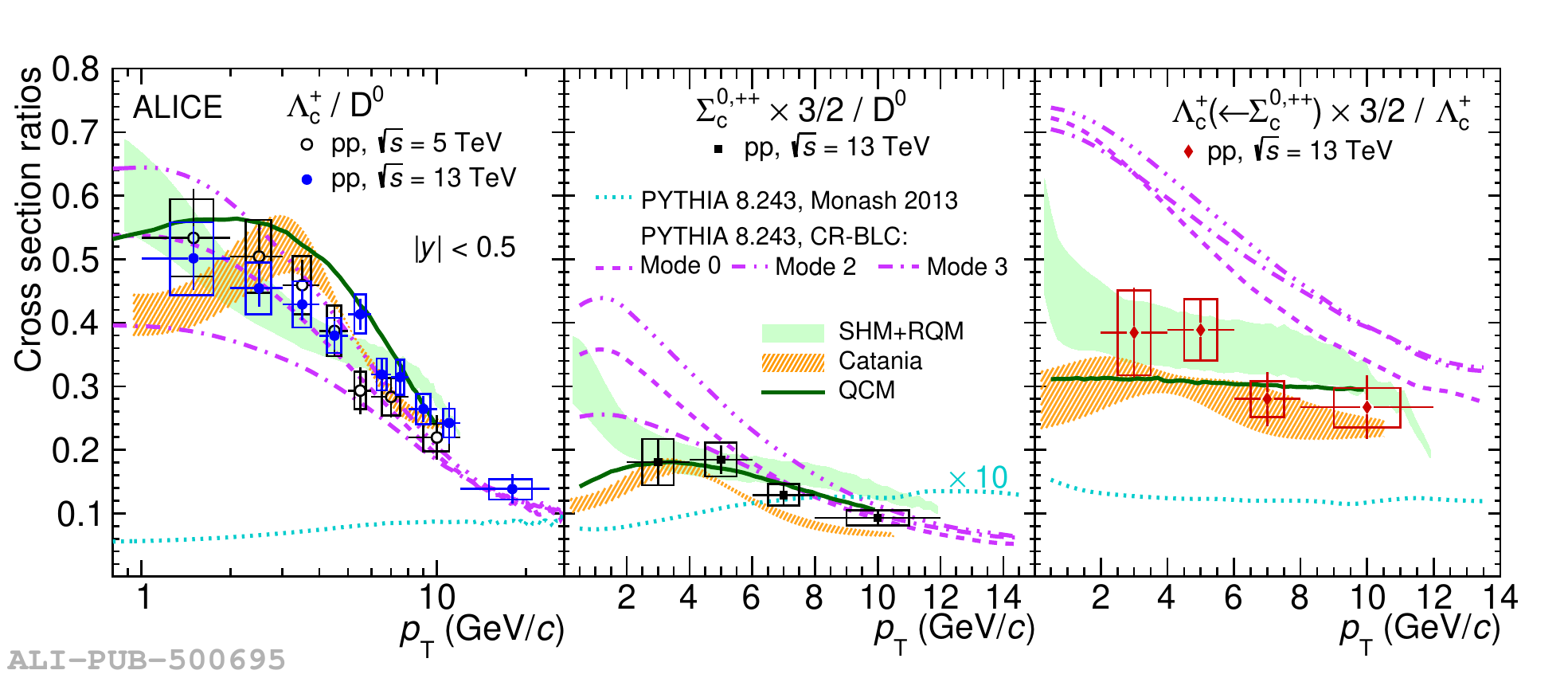}
    \caption{Measurements of (left) \LcD, (middle) \Sigmac/\Dz, and (right) contribution of \Sigmac feed-down to the prompt \Lc cross section in pp collisions at \sqrtsthirt \cite{ALICE:2021rzj}. The measurements are compared to model calculations~\cite{He:2019tik,Christiansen:2015yqa,Skands:2014pea,Minissale:2020bif,Li:2017zuj}.}
    \label{fig:Sigmac}
\end{figure}

Measurements of the $\Sigmac/\Dz$ yield ratio~\cite{ALICE:2021rzj} in \figref{fig:Sigmac} (middle) also show a significant enhancement, by a factor larger than 10, with respect to the Monash tune of PYTHIA~\cite{Skands:2014pea}, while it is well described by the statistical hadronization model~\cite{He:2019tik}, and models including coalescence~\cite{Minissale:2020bif,Li:2017zuj}. The $\Sigmac$ baryon states are contributors to the $\Lc$ yield as they decay strongly and are not distinguishable from prompt production. The feed-down contribution $\Lc(\leftarrow \Sigmac)$ is shown in the right panel of \figref{fig:Sigmac} for \pt$>2$~GeV/c, and the \pt integrated fraction is found to be $0.38\pm 0.06 \mathrm{(stat.)} \pm 0.06 \mathrm{(syst.)}$.
A suppression of these states with respect to the $\Lc$ baryon production is expected due to the relative rarity of spin-1 diquarks. This implies that a sizable proportion of the measured prompt $\Lc$ originates from the decays of heavier resonant states, and notably the PYTHIA predictions with enhanced color reconnection appear to overestimate this contribution.

\subsection{\boldmath{\xicz}, \boldmath{\xicp} and \boldmath{\Omegac} production in pp collisions at \boldmath{\sqrtsthirt}}

\begin{figure}[h!tb]
    \centering
    \includegraphics[width=0.53\textwidth]{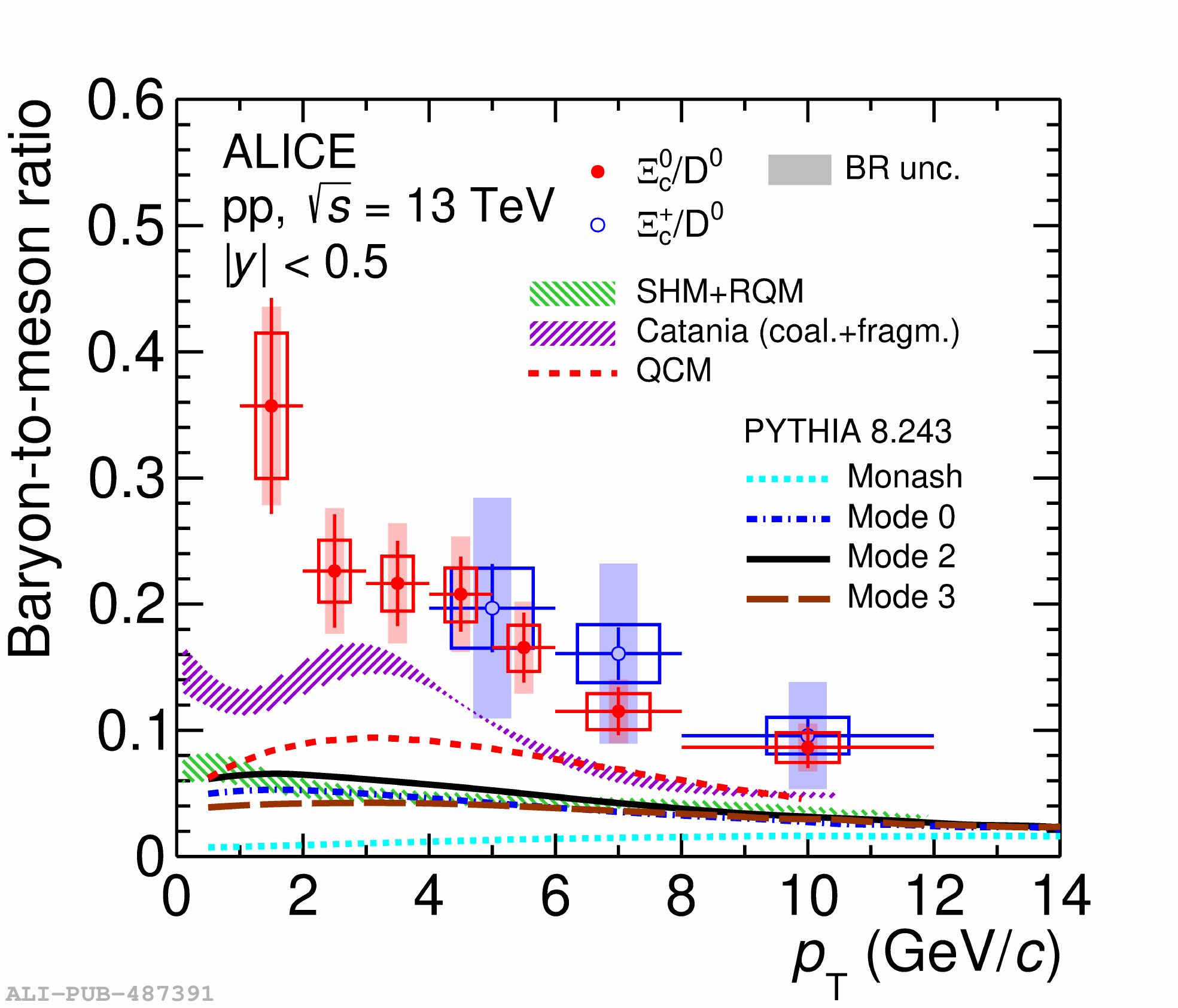}
    \includegraphics[width=0.46\textwidth]{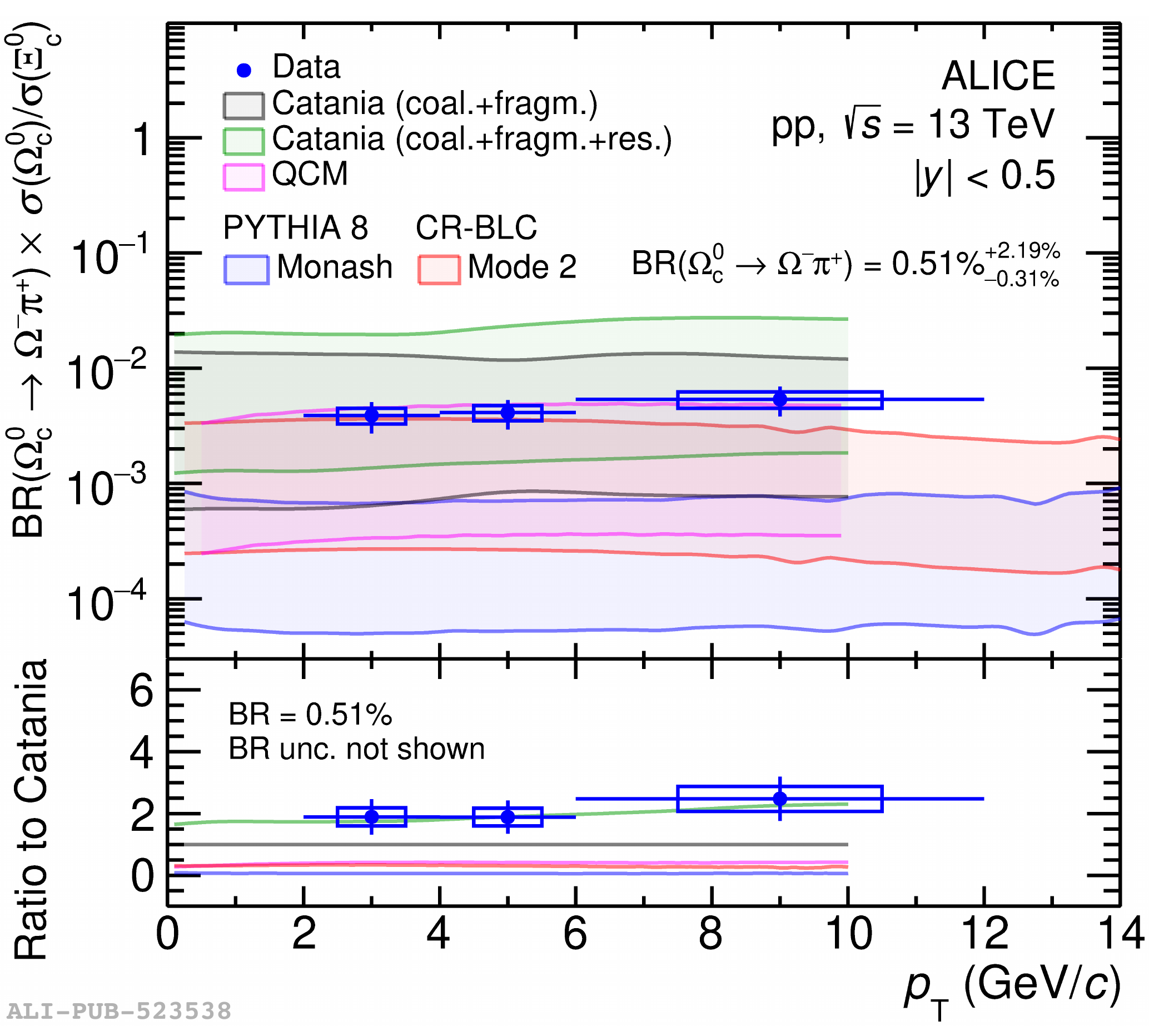}
    \caption{Left: Measurement of $\xicz/\Dz$ and $\xicp/\Dz~$\cite{ALICE:2021bli} in comparison to models~\cite{He:2019tik,Christiansen:2015yqa,Skands:2014pea,Minissale:2020bif,Li:2017zuj}. Right: Baryon-to-baryon yield ratio $\mathrm{BR}(\Omegac \rightarrow \Omega^{-} \pi^{+})\times\sigma(\Omegac)/\sigma(\xicz)$ in pp collisions at \sqrtsthirt~\cite{Omegac}. } 
    \label{fig:OcXic}
\end{figure}

The impact of strangeness on the hadronization process can be studied by measuring $\xiczp$~\cite{ALICE:2021bli} and $\Omegac$~\cite{Omegac} baryons. The measurements were performed in pp collisions at \sqrtsthirt. The $\xicz/\Dz$ and $\xicp/\Dz$ yield ratios are displayed in the left panel of \figref{fig:OcXic}, and the right panel shows the baryon-to-baryon yield ratio $\Omegac/\xicz$. Since the branching ratio of the decay $\Omegac \rightarrow \Omega^{-} \pi^{+}$ is not yet experimentally measured, the $\Omegac$ cross section is not corrected for the theoretical branching ratio~\cite{hsiao2020charmed}. The measurements show a significant enhancement over all models, even those that were able to describe the \LcD yield ratio. The Catania model~\cite{Minissale:2020bif} comes closest to describing the data. PYTHIA with Monash tune~\cite{Skands:2014pea} underestimates the data by orders of magnitude and even PYTHIA including CR beyond leading color approximation~\cite{Christiansen:2015yqa} underestimates the data. The Catania model can also describe the data when including higher mass resonance decays. Using the ALICE \xicz data and the $\Omegac/\xicz$ yield ratio, scaled by the branching ratio, the integrated \Omegac cross section can be obtained. Compared to measurements from the Belle Collaboration in \ee collisions at $\sqrt{s} = 10.52~\TeV$~\cite{PhysRevD.97.072005}, the ratio estimated by ALICE is larger by a factor of $4.7 \pm 1.3\mathrm{(stat.)} \pm 0.5\mathrm{(syst.)}$ for the $\mathrm{BR}(\Omegac \rightarrow \Omega^{-} \pi^{+})\times\sigma(\Omegac)/\sigma(\xicz)$ yield ratio. 

\subsection{Total charm cross section and fragmentation fractions}

\begin{figure}[h!tb]
    \centering
    \includegraphics[width=0.49\textwidth]{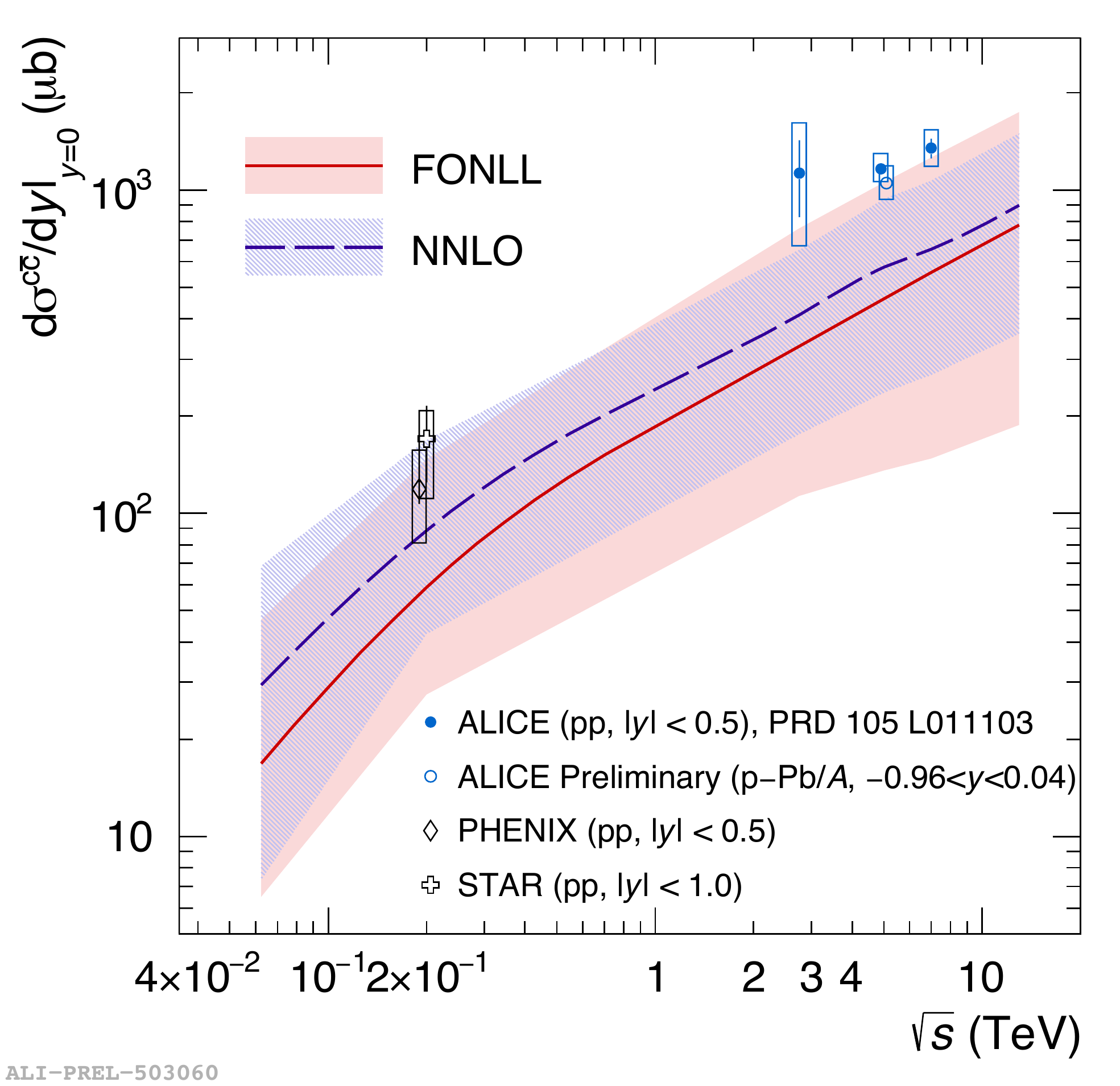} 
    \includegraphics[width=0.49\textwidth]{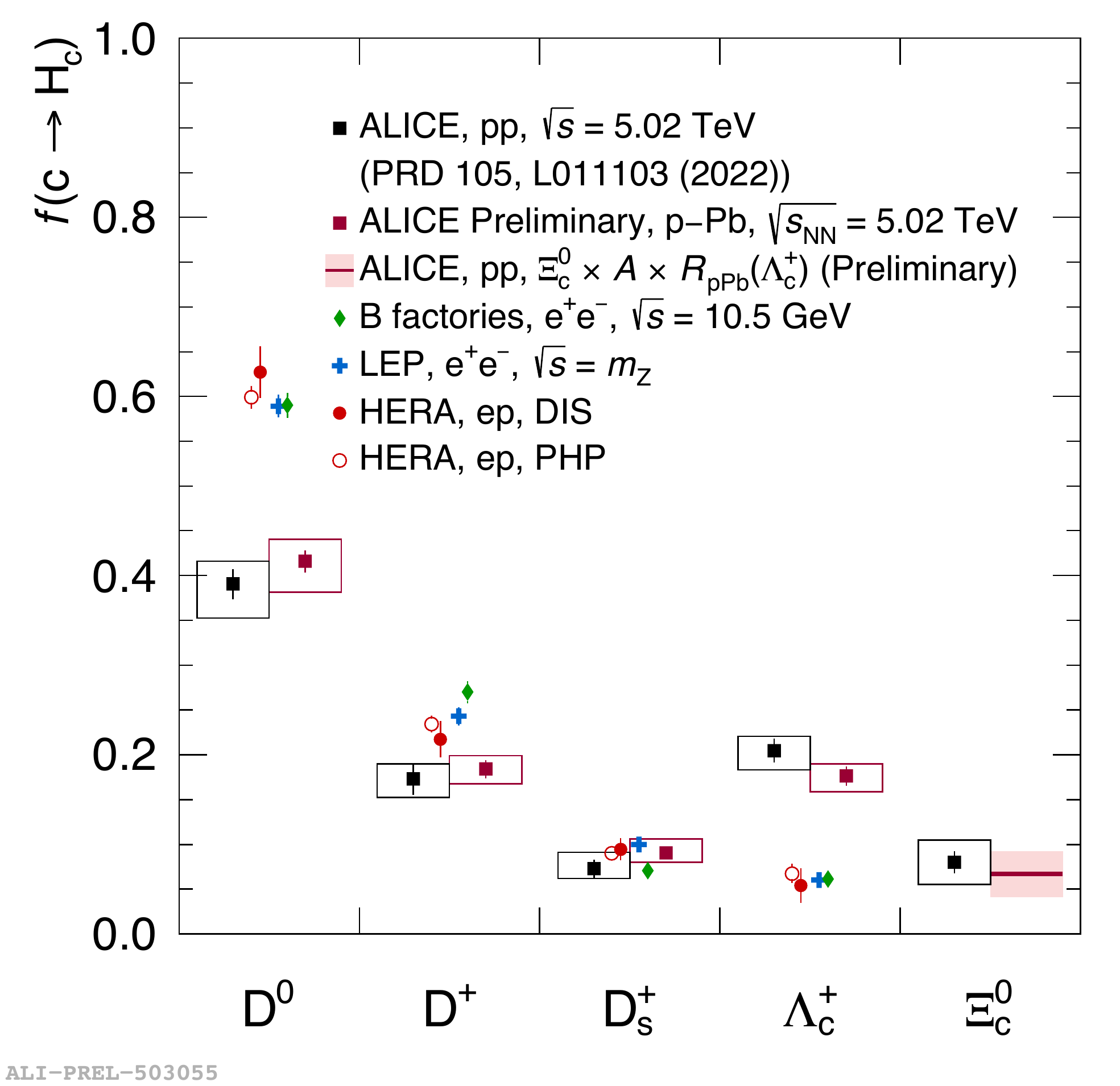}

    \caption{Left: Total charm production cross section at midrapidity in pp and \pPb collisions as a function of collision energy~\cite{acharya2022charm} in comparison to FONLL and NNLO calculations~\cite{Cacciari:1998it}. Right: Relative hadronization fractions of ground-state charm hadron species measured by ALICE in pp and \pPb collisions, compared with \ee and \ep collisions~\cite{lisovyi2016combined}.} 
    \label{fig:ccvsenergy}
\end{figure}

The total charm cross section per unit of rapidity at mid rapidity is computed as the sum of the cross section of the single charm ground state hadron species, and shown in \figref{fig:ccvsenergy} (left) as a function of the collision energy. The open blue marker shows the first computation of the total $\mathrm{c\bar{c}}$ cross section in \pPb collisions at \sqrtsNNfive, and it is shown together with previous ALICE measurements in pp collisions at $\sqrt{s} = 2.76, 5.02$ and $7~\TeV$~\cite{acharya2022charm}, shown as full blue markers. A comparison with FONLL and NNLO calculations shows that the ALICE results lie on the upper edge of the uncertainty bands of the pQCD calculations. In the right panel of \figref{fig:ccvsenergy}, the individual fragmentation fractions of the different hadron species are shown. The results in pp collisions are shown together with the preliminary measurements in \pPb collisions. The first measurement of the \xicz in pp collisions at \sqrtsfive is also shown. The measurements are compared to \ee and \ep collisions~\cite{lisovyi2016combined}, and an enhancement of \Lc baryon production can be seen with respect to leptonic collisions, which corresponds to a depletion of the meson abundances.

\section{Outlook}
In the upcoming Run 3 and 4 and future ALICE 3 experiment~\cite{ALICE3LoI} many more exciting results on heavy flavor production and hadronization are expected. An upgraded apparatus and a higher data taking rate allows to perform measurements with even more precision in a broader momentum range. Furthermore, the full reconstruction of beauty hadrons and multi-charm species will be feasible. 

\section{Bibliography}
\bibliography{bibliography.bib}
\end{document}